\documentclass[10pt,a4paper,oneside]{IEEEtran}
\usepackage{silence}
\newlength{\flexwidth}
\usepackage{bm}
\usepackage{amsmath,amssymb,amsthm,fixmath}
\usepackage{mathtools}
\usepackage{accents}
\usepackage{color,colortbl}
\usepackage{xcolor}
\usepackage{siunitx}
\usepackage{multirow,booktabs}
\usepackage{subcaption}
\usepackage[inline]{enumitem}
\usepackage{arydshln}
\usepackage{optidef}
\usepackage{graphicx}
\usepackage{lipsum}

\usepackage{amsmath}
\usepackage{epstopdf}
\usepackage[normalem]{ulem} 
\usepackage{xcolor}

\usepackage[normalem]{ulem}

\usepackage[textsize=tiny,colorinlistoftodos]{todonotes}
\makeatletter
\define@key{todonotes}{bh}[]{
	\setkeys{todonotes}{author=\textbf{Bin}, color=lime!30}}%
\define@key{todonotes}{zf}[]{
	\setkeys{todonotes}{author=\textbf{Zexin}, color=blue!30}}%
\makeatother

\newif\ifreviewmode
\reviewmodetrue 

\ifreviewmode
\else
  \renewcommand{\todo}[1]{} 
\fi

\usepackage[shortcuts,acronym,automake]{glossaries}
\makeglossaries
\newacronym{ols}{OLS}{Ordinary Least Squares}
\newacronym{lmm}{LMM}{Linear Mixed Model}
\newacronym{mir}{MIR}{music information retrieval}
\newacronym{sr}{SR}{stochastically regulated}
\newacronym{fm}{FM}{frequency modulated}
\newacronym{rms}{RMS}{Root Mean Square}
\newacronym{asmr}{ASMR}{Autonomous Sensory Meridian Response}
\newacronym{scd}{SCD}{Spectral Correlation Density}
\newacronym{ccf}{CCF}{Cyclic Coherence Function}
\newacronym{eeg}{EEG}{electroencephalography}
\newacronym{gan}{GAN}{Generative Adversarial Network}
\newacronym{ai}{AI}{Artificial Intelligence}

\usepackage{tcolorbox}
\tcbuselibrary{many}

\newtcbtheorem[]{alg}{Algorithm}{fonttitle=\bfseries}{alg}

\usepackage[norelsize,linesnumbered,ruled]{algorithm2e}
\SetKwRepeat{Do}{do}{while}
\makeatletter
\newcommand{\removelatexerror} {\let\@latex@error\@gobble}
\makeatother
\begin{document}
	\title{Is ASMR Engineerable? A Signal Processing and User Experience Study}	
	\author{{Zexin~Fang},~\IEEEmembership{Student Member,~IEEE,}
		{Bin~Han},~\IEEEmembership{Senior Member,~IEEE,}
		{Henrik~H.~Sveen},
        {C.~Clark~Cao}, 
        and
		{Hans~D.~Schotten},~\IEEEmembership{Member,~IEEE}
		 \thanks{Z. Fang, B. Han, and H. D. Schotten are with {University of Kaiserslautern-Landau (RPTU), Germany}. H. D. Schotten is with the German Research Center for Artificial Intelligence (DFKI), Germany. {C. Clark Cao is with Lingnan University, Hong Kong. Henrik. H. Sveen is with University of Oslo, Norway. B. Han (bin.han@rptu.de) is the corresponding author. This work is supported by the German Federal Ministry of Education and Research within the project Open6GHub (16KISK003K/16KISK004). }}
	}\maketitle
	\begin{abstract}  
       \gls{asmr} has been remarkably popular in the recent decade, yet whether its effects can be deliberately engineered remains an open question. While \gls{asmr} effects validated through behavioral studies and neuro-physiological measurements such as \gls{eeg} and related bio-signals, the acoustic mechanisms that trigger it remain poorly understood. We investigate whether \gls{asmr} responses can be systematically induced through controlled acoustic design, hypothesizing that cyclic patterns where predictability drives relaxation and variation sustains intrigue are key engineerable parameters. Specifically, we design cyclic sound patterns with varying predictability and randomness, and evaluate their effects via a structured user study. Signal processing-based feature extraction and regression analysis are used to establish an interpretable mapping between acoustic structure and perceived \gls{asmr} effects. Results show that relaxing effects accumulate progressively, are independent of spatial orientation, and remain stable across time. Crucially, smoothly spread, energy-dense cyclic patterns most effectively trigger \gls{asmr}, suggesting that signal-level engineering of \gls{asmr} experiences is achievable.
	\end{abstract}
    \glsresetall

 \section{Introduction}
  \gls{asmr} has gained significant attention as a unique sensory experience capable of inducing relaxation and engagement through audiovisual stimuli. Its growing use in digital media highlights its potential for human-centered applications such as immersive content and affective computing. However, despite its popularity, it remains unclear whether \gls{asmr} responses can be systematically engineered through controllable acoustic design.


Beyond its popularity, \gls{asmr} has attracted considerable academic interest. Prior studies have shown that \gls{asmr} modulates physiological responses (e.g., heart rate and skin conductance) and is associated with individual differences in emotional sensitivity \cite{poerio2018more, fredborg2017asmr, eid2022untangling}. Neuroimaging and behavioral studies further suggest that \gls{asmr} involves atypical interactions between auditory processing and affective systems, similar to related phenomena such as misophonia \cite{lochte2018fMRI, smith2019fMRI}. EEG-based analyses additionally indicate that audio frequency modulates brain activity across multiple bands ($\theta
,\alpha, \beta, \gamma$ and high $\gamma$), with \gls{asmr} responders maintaining higher mental engagement while exhibiting central activity suppression \cite{liasmr2024}.

However, existing studies primarily address perception and physiological responses, while the acoustic principles underlying controllable and reproducible \gls{asmr} generation remain poorly understood. \gls{asmr} effects share similarities with musical frisson \cite{del2016autonomous}. Analyses of Zwicker parameters (loudness, sharpness, roughness, and fluctuation strength) and ACF/IACF suggest that nature-generated sounds with higher loudness and roughness, and human-generated sounds with lower IACC, are more likely to evoke \gls{asmr} sensations \cite{shimokura2022sound}. Other studies highlight that semantic content may contribute to broader social-affective processes such as relaxation and prosocial bonding \cite{villena2023asmr, sakurai2023asmr}. However, these studies rely on naturalistic clips rich in semantic information, where cognitive biases and personal preferences are deeply entangled with acoustic features. Can acoustic structure alone, stripped of semantic content, reliably engineer \gls{asmr} experiences?
In our previous work, \gls{gan}-generated audio trained solely on acoustic patterns effectively induced \gls{asmr} responses, suggesting that cyclic patterns are a key engineerable feature \cite{ZexinASMR2023}. Research further suggests predictability and novelty as two governing factors of \gls{asmr} triggering \cite{barratt2017sensory}, which can be interpreted as the degree of cyclic regularity and variation within it.

  Inspired by prior research, this paper quantitatively investigates how predictability and randomness in acoustic features influence \gls{asmr} perception, combining a behavioral study with audio signal processing analysis. We designed synthetic \gls{asmr} stimuli with controlled predictability and randomness to ensure diverse statistical properties and discriminable \gls{asmr} response scores (Sec.~\ref{sec:audiogenerate}). Cyclostationary analysis was then applied to extract statistical features from the stimuli (Sec.~\ref{sec:feature}). \gls{asmr} response data were collected via an online questionnaire in which participants rated their subjective experience for each stimulus (Sec.~\ref{sec:survey}). Finally, regression analysis was performed to establish an explainable mapping between the extracted acoustic features and \gls{asmr} responses, with results and discussion presented in Sec.~\ref{sec:results}, and conclusions drawn in Sec.~\ref{sec:conclu}.
   
 \section{Audio Clips Generation}\label{sec:audiogenerate}
   To investigate the relationship between cyclic patterns and \gls{asmr} response, we employ three distinct types of patterns: \gls{mir} cyclic patterns, \gls{sr} cyclic patterns, and \gls{fm} cyclic patterns. Since \gls{ai}-generated clips offer limited control over internal acoustic structure, synthetic stimuli are adopted to enable precise and reproducible manipulation of predictability and randomness. \gls{mir} cyclic patterns directly resemble the interaction of an \gls{asmr} artist, with cyclic patterns extracted from highly popular nail-tap \gls{asmr} clips on open-access platforms. For \gls{sr} patterns, predictability is addressed by repeating a bundle of impulses every \SI{1.6}{\second} to mimic highly popular \gls{asmr} patterns, while randomness is introduced by stochastically adjusting the impulse density within each bundle, with a minimum impulse duration of \SI{40}{\milli\second} in accordance with the precedence effect \cite{haas1972influence}. Similarly, \gls{fm} cyclic patterns repeat every \SI{1.6}{\second}, with impulse density varying linearly from \SI{0}{\hertz} to \SI{25}{\hertz}. Given the role of immersion in \gls{asmr} responses, both monophonic and stereophonic spatial orientations are investigated, with the stereophonic condition extracted from nail-tap \gls{asmr} clips to preserve a connection with real-world scenarios.
   \begin{figure}[ht]
    \centering
    \begin{subfigure}[b]{0.36\textwidth}
        \centering\caption{Spectrogram of white noise}
        \includegraphics[width=\textwidth]{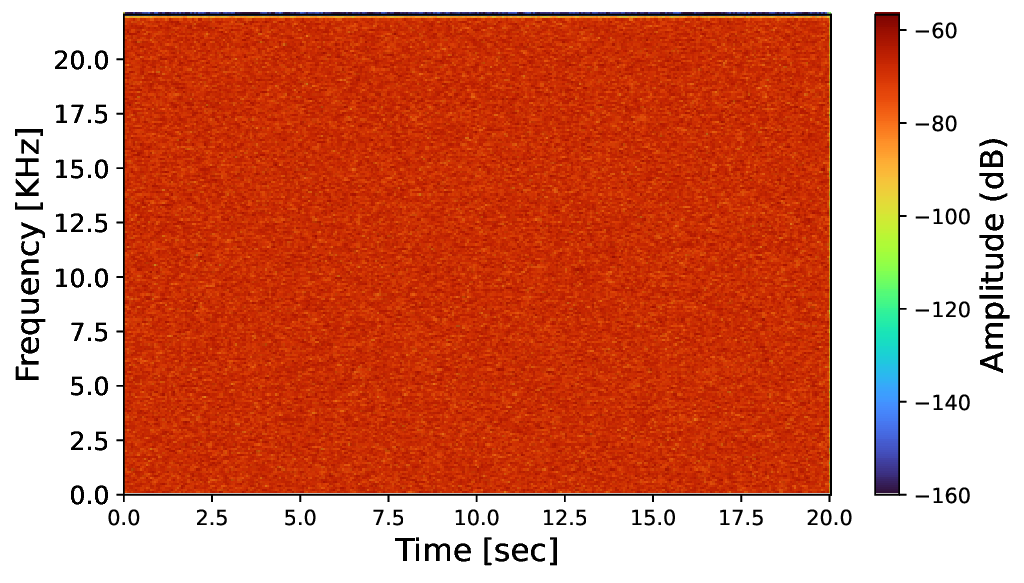}
    \end{subfigure}
    \hfill
    \begin{subfigure}[b]{0.36\textwidth}
        \centering\caption{Spectrogram of MIR clip}
        \includegraphics[width=\textwidth]{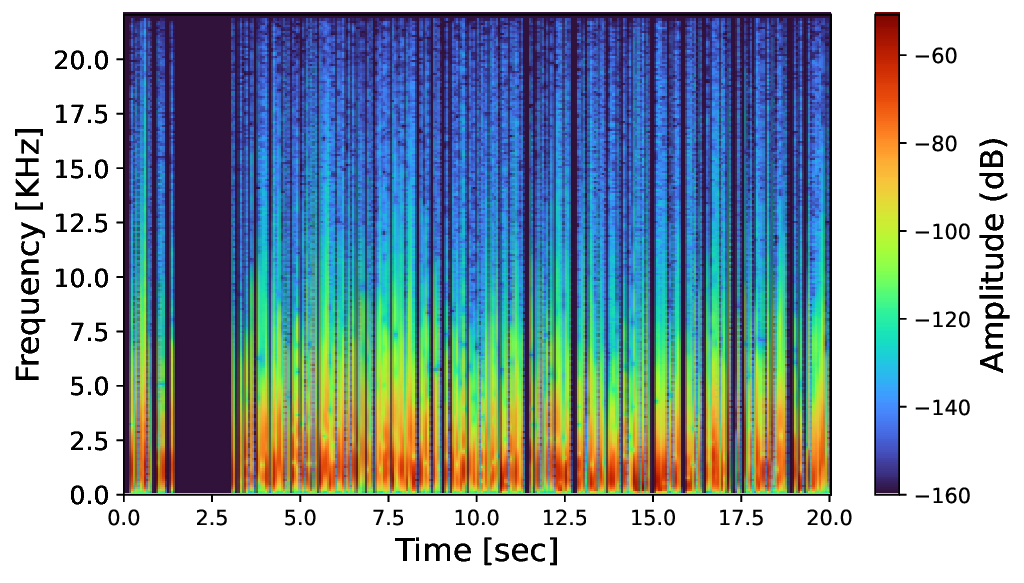}
    \end{subfigure}
    \begin{subfigure}[b]{0.36\textwidth}
        \centering\caption{Spectrogram of SR clip}
        \includegraphics[width=\textwidth]{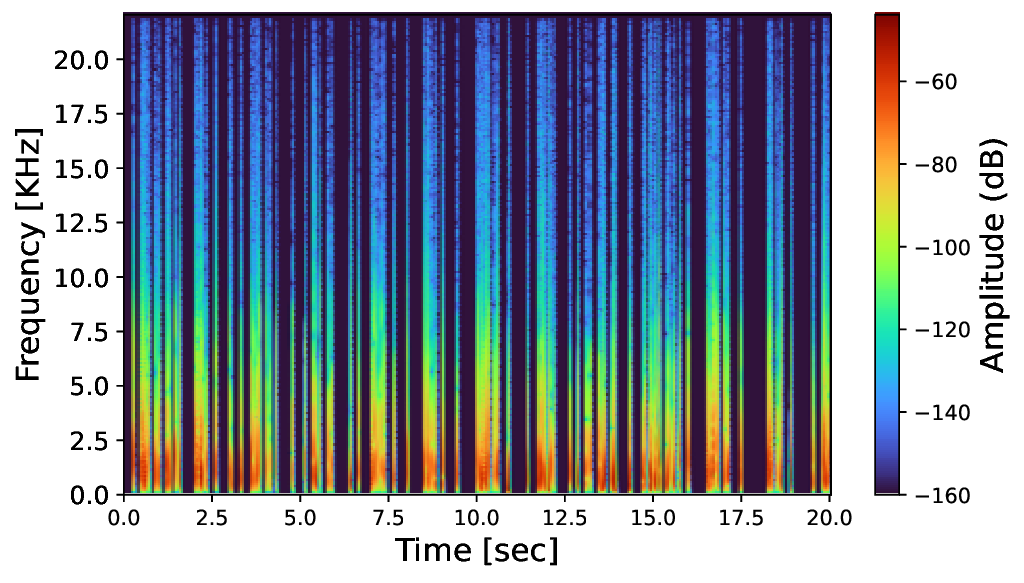}
    \end{subfigure}
    \begin{subfigure}[b]{0.36\textwidth}
        \centering\caption{Spectrogram of FM clip}
        \includegraphics[width=\textwidth]{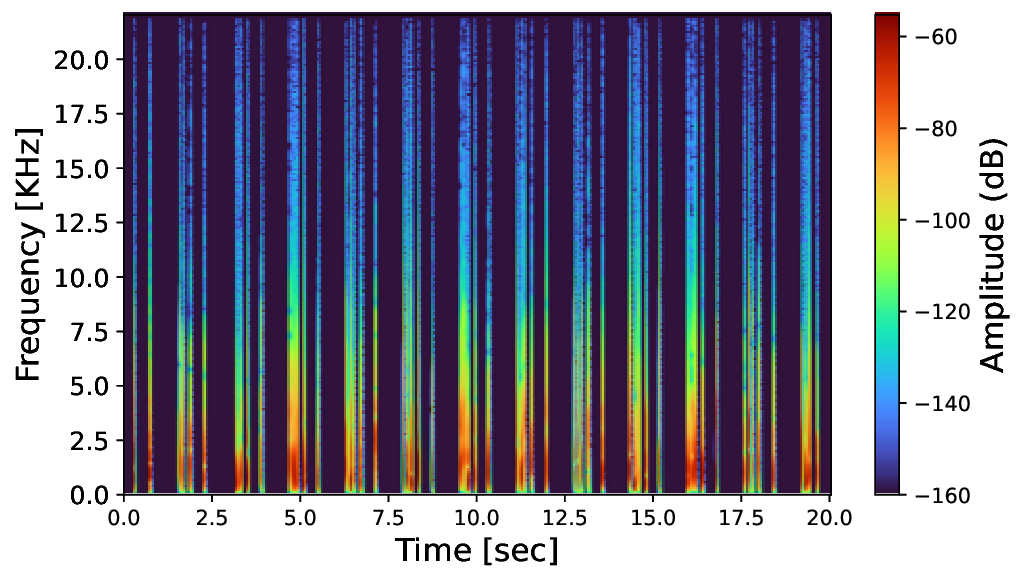}
    \end{subfigure}
    \caption{Spectrograms of different audio clips}
   \end{figure}
 \section{Statistical Features Extraction}\label{sec:feature}
 
 For a discrete signal $s(n)$ containing $N$ samples, the spectrum $S(f)$ can be computed through the Discrete Fourier Transform (DFT):
\begin{equation}
	S(f)=\frac{1}{N}\sum\limits_{k=0}^{N-1}\left[\sum\limits_{n=0}^{N-1}s(n)e^{-\frac{j2\pi nk}{N}}\delta\left(2\pi f-\frac{2\pi k}{N}\right)\right]. \label{eq:spect}
\end{equation}

 To characterize cyclic features within the Eq.~(\ref{eq:spect}), we employ the \gls{scd} fuction, with
 \begin{align}
	D_{SS}(f,\alpha) &= \mathbb{E}\left\{S(f+\frac{\alpha}{2})S^{*}(f-\frac{\alpha}{2})\right\},
 \end{align}
where $\alpha$ represents the cyclic frequency, indicating spectral periodicity. While \gls{scd} is inherently a bivariate function, we introduce two univariate functions of $\alpha$ to capture the cyclic characteristics of an audio signal $s(n)$. These are defined as $\overline{D}_{SS}(\alpha) = \frac{1}{N} \sum\limits_{f\in\mathbb{F}} D_{SS}(f,\alpha)$ and $\Tilde{D}_{SS}(\alpha) = \max\limits_{f\in\mathbb{F}} D_{SS}(f,\alpha)$, where $\mathbb{F}$ represents the set of $N$ frequency bins in $D_{SS}$.

\begin{figure} 
    \centering
    \begin{subfigure}[b]{0.241\textwidth}
        \centering
        \includegraphics[width=\textwidth]{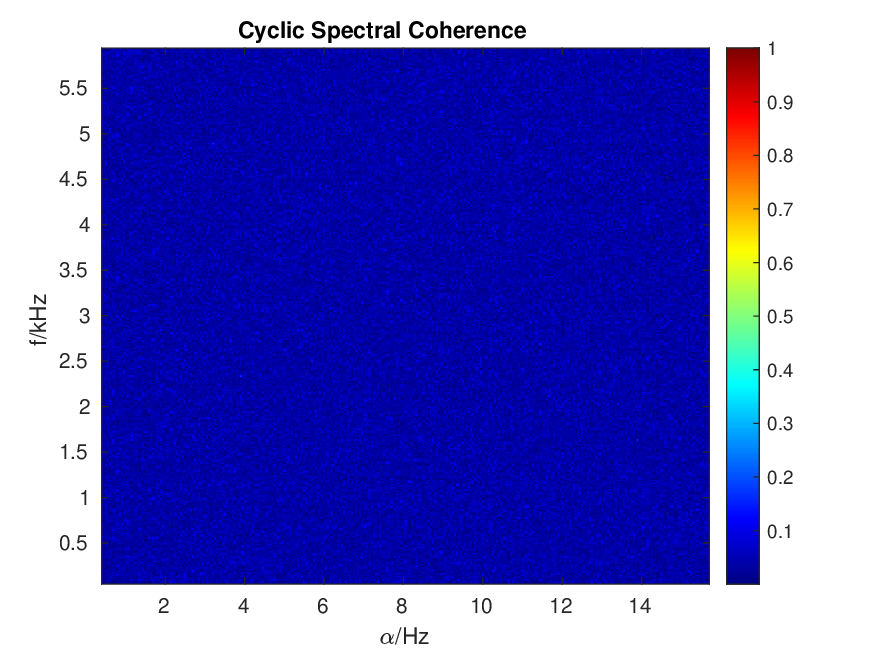}
        \caption{white noise}
    \end{subfigure}
    \hfill
    \begin{subfigure}[b]{0.241\textwidth}
        \centering
        \includegraphics[width=\textwidth]{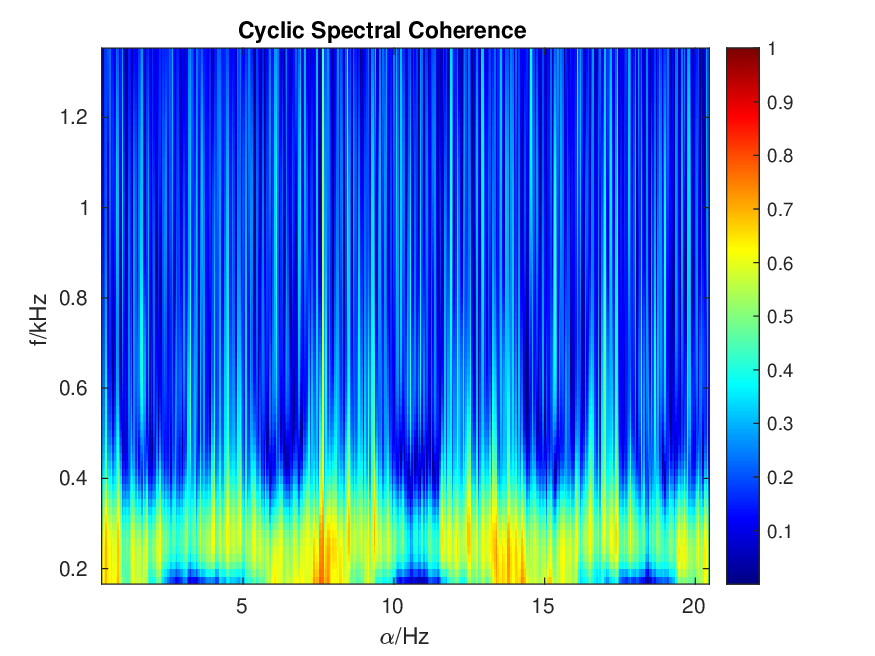}
        \caption{\gls{mir} cyclic patterns}
    \end{subfigure}
    \begin{subfigure}[b]{0.241\textwidth}
        \centering
        \includegraphics[width=\textwidth]{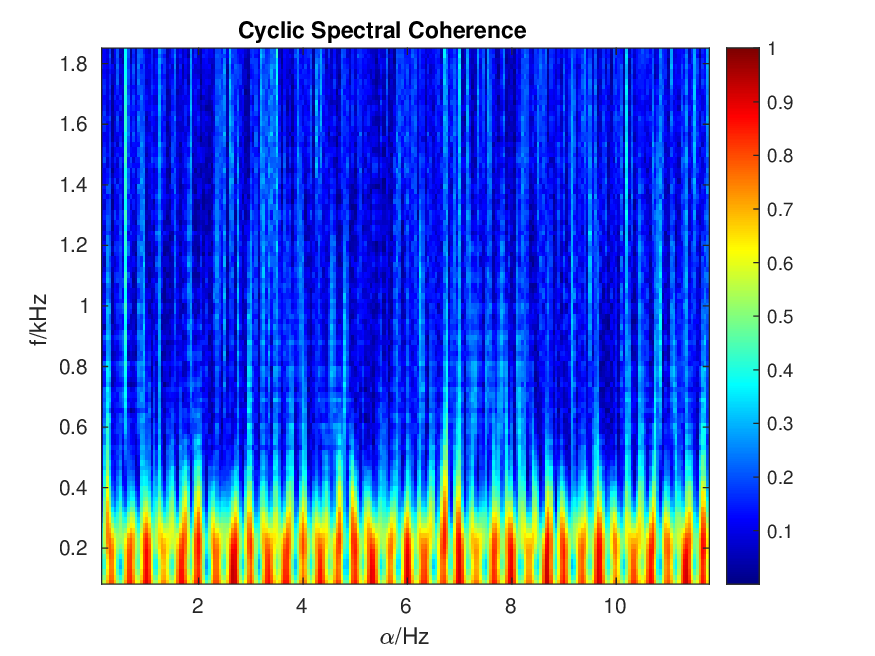}
        \caption{\gls{sr} cyclic patterns}
    \end{subfigure}
    \begin{subfigure}[b]{0.241\textwidth}
        \centering
        \includegraphics[width=\textwidth]{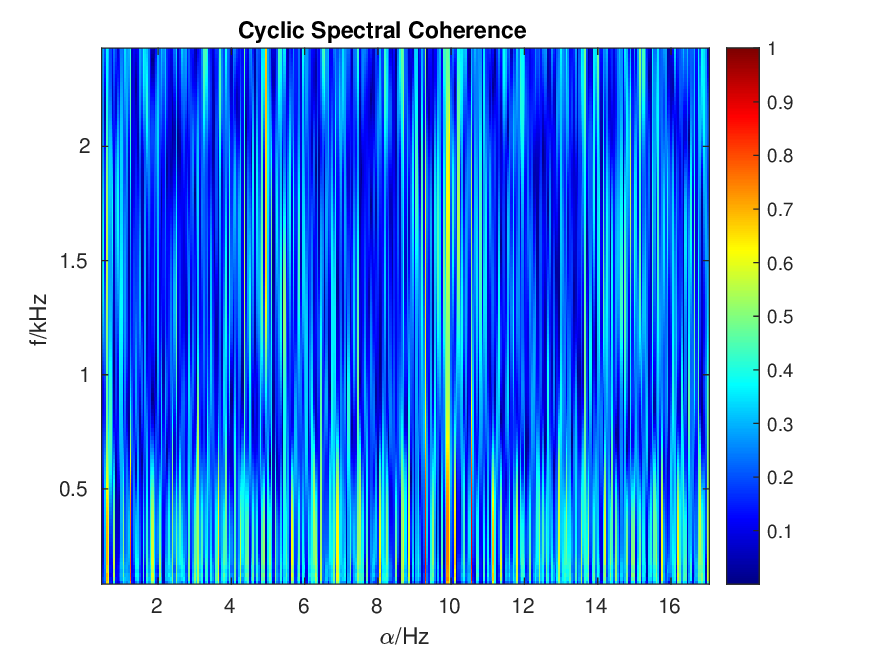}
        \caption{\gls{fm} cyclic patterns}
    \end{subfigure}
    \caption{\gls{ccf} of \gls{asmr} audio clips with different cyclic patterns}
    \label{fig:ccf_all}
   \end{figure}

\begin{figure} 
    \centering
    \begin{subfigure}[b]{0.241\textwidth}
        \centering
        \includegraphics[width=\textwidth]{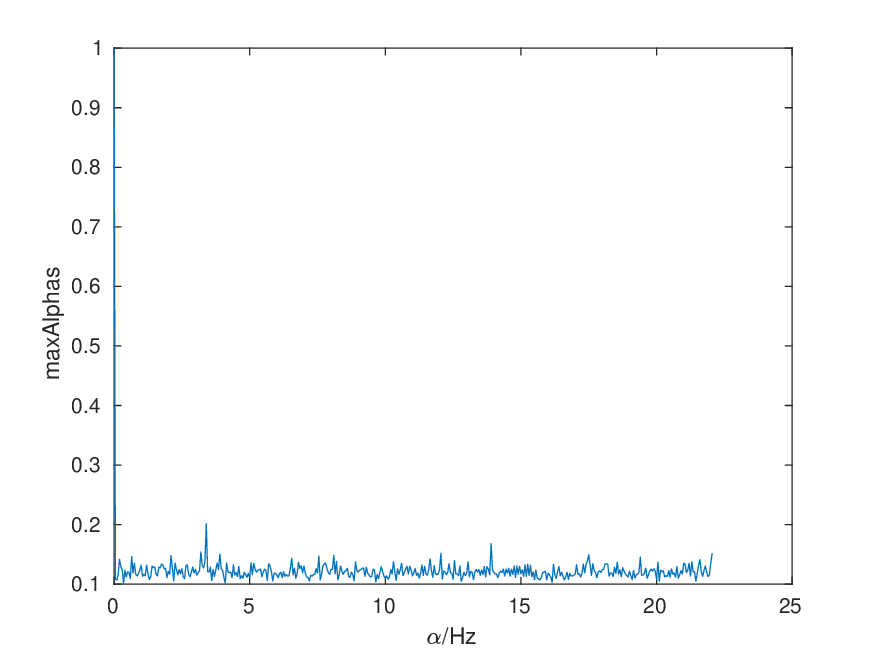}
        \caption{white noise}
    \end{subfigure}
    \hfill
    \begin{subfigure}[b]{0.241\textwidth}
        \centering
        \includegraphics[width=\textwidth]{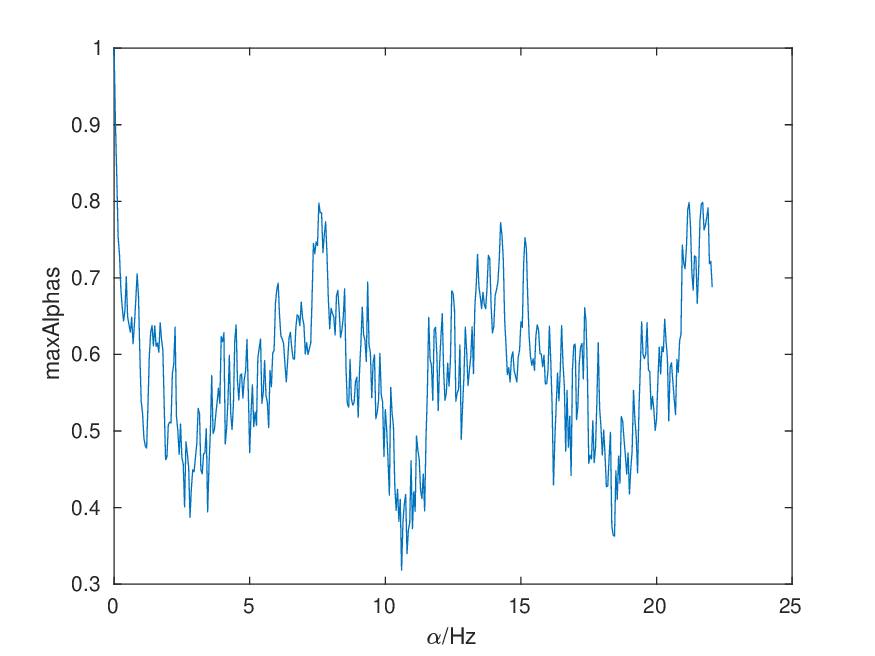}
        \caption{\gls{mir} cyclic patterns}
    \end{subfigure}
    \begin{subfigure}[b]{0.241\textwidth}
        \centering
        \includegraphics[width=\textwidth]{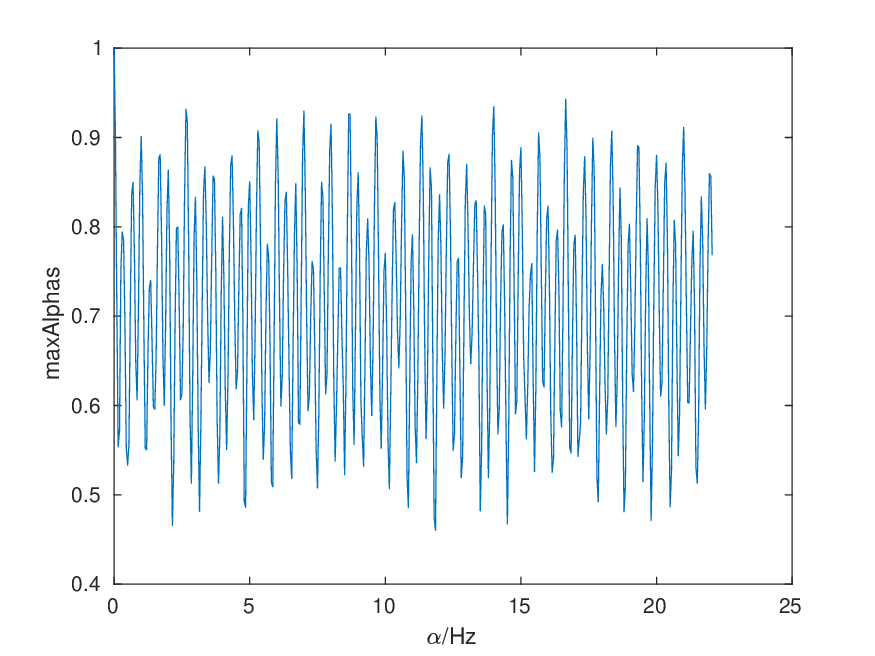}
        \caption{\gls{sr} cyclic patterns}
    \end{subfigure}
    \begin{subfigure}[b]{0.241\textwidth}
        \centering
        \includegraphics[width=\textwidth]{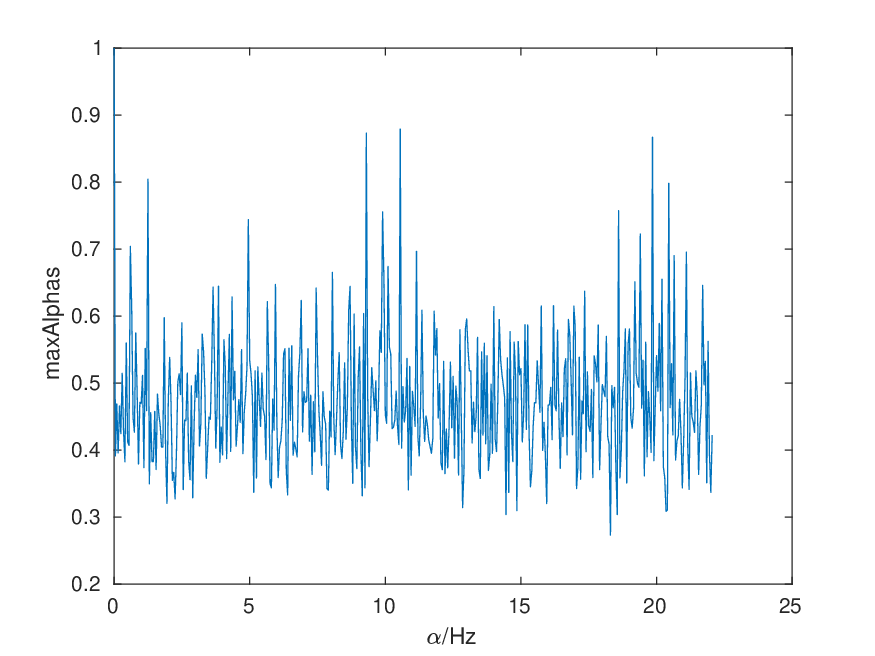}
        \caption{\gls{fm} cyclic patterns}
    \end{subfigure}
    \caption{$\Tilde{D}_{SS}(\alpha)$ of \gls{asmr} audio clips with different cyclic patterns}
    \label{fig:amax_all}
   \end{figure}

\begin{figure} 
    \centering
    \begin{subfigure}[b]{0.241\textwidth}
        \centering
        \includegraphics[width=\textwidth]{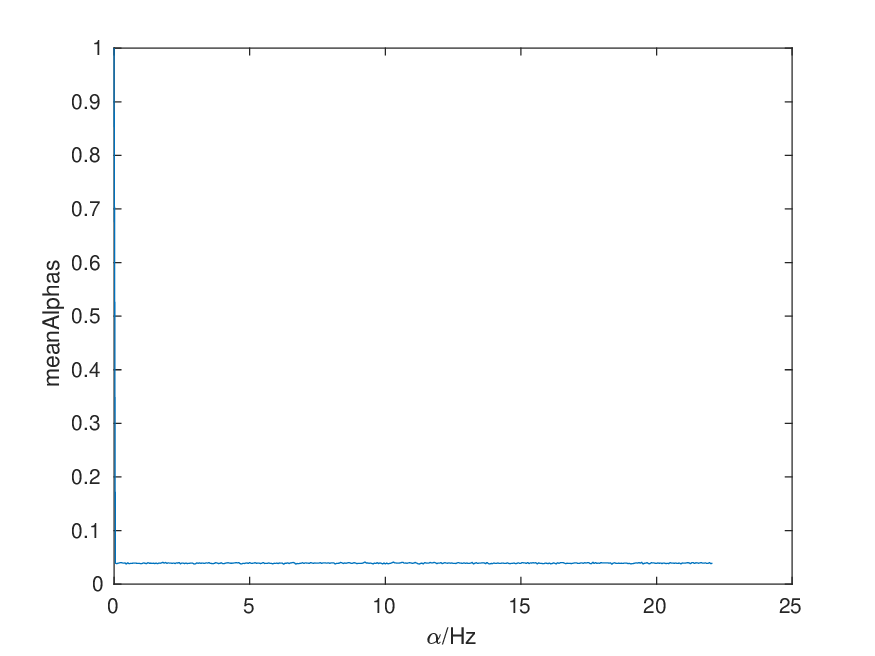}
        \caption{white noise}
    \end{subfigure}
    \hfill
    \begin{subfigure}[b]{0.241\textwidth}
        \centering
        \includegraphics[width=\textwidth]{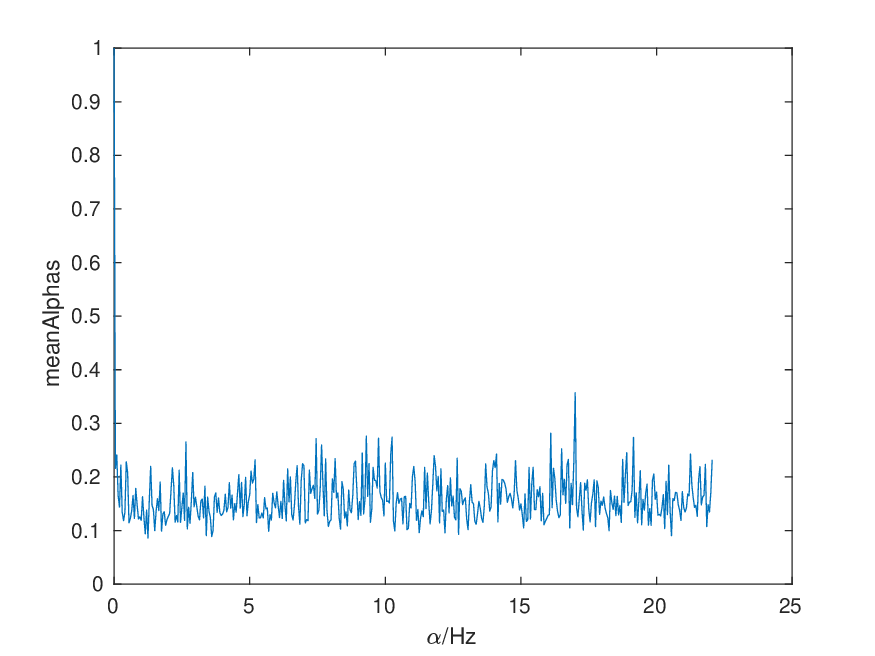}
        \caption{\gls{mir} cyclic patterns}
    \end{subfigure}
    \begin{subfigure}[b]{0.241\textwidth}
        \centering
        \includegraphics[width=\textwidth]{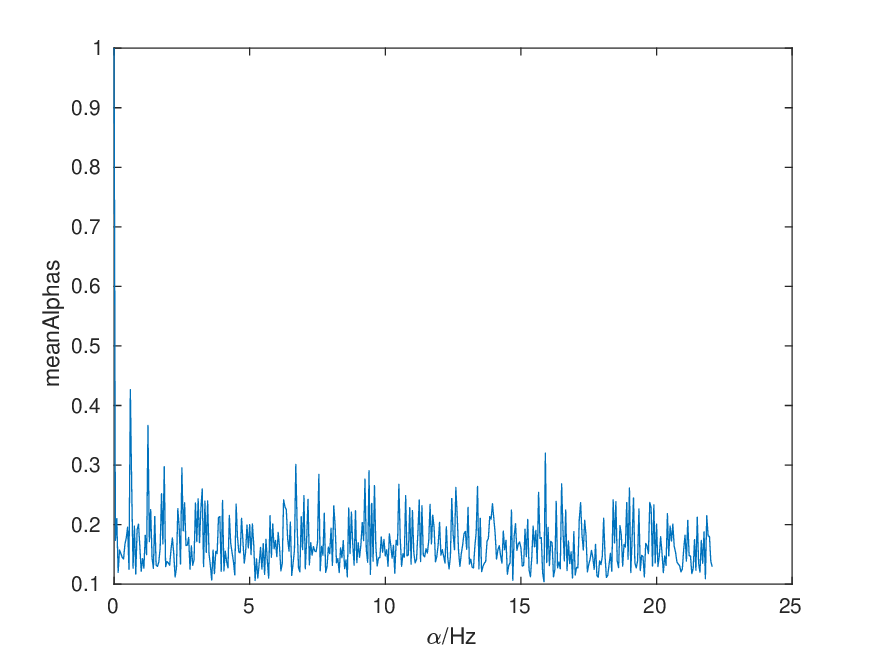}
        \caption{\gls{sr} cyclic patterns}
    \end{subfigure}
    \begin{subfigure}[b]{0.241\textwidth}
        \centering
        \includegraphics[width=\textwidth]{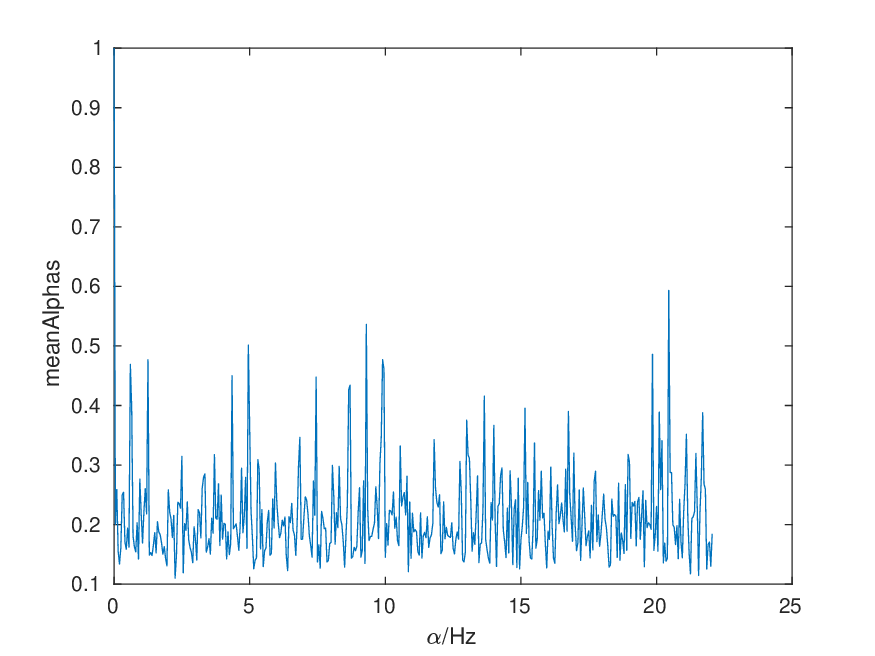}
        \caption{\gls{fm} cyclic patterns}
    \end{subfigure}
    \caption{$\overline{D}_{SS}(\alpha)$ of \gls{asmr} audio clips with different cyclic patterns}
    \label{fig:amean_all}
   \end{figure}

On top of $S(f)$, $\Tilde{D}_{SS}(\alpha)$ and $\overline{D}_{SS}(\alpha)$ , we define eight features: $X_1={\mathrm{Mean}}\left(\Tilde{D}_{SS}\right)$, $X_2={\mathrm{Var}}\left(\Tilde{D}_{SS}\right)$, $X_3={G}\left(\Tilde{D}_{SS}\right)$, $X_4={\mathrm{Mean}}\left(\overline{D}_{SS}\right)$, $X_5={\mathrm{Var}}\left(\overline{D}_{SS}\right)$, $X_6={G}\left(\overline{D}_{SS}\right)$, $X_7={\mathrm{Mean}}\left(\left\vert S\right\vert\right)$, 
$X_8={\mathrm{Var}}\left(\left\vert S\right\vert\right)$, and $X_9={G}\left(\left\vert S\right\vert\right)$, where $G(\mathbb{S})=\frac{\sum\limits_{i\in\mathbb{S}}\sum\limits_{j\in\mathbb{S}}\vert i-j \vert}{2\Vert\mathbb{S}\Vert_0\sum\limits_{i\in\mathbb{S}} i}$ is the Gini coefficient of set $\mathbb{S}$, which was originally used to describe income inequality. Here, $X_1$ and $X_4$ assess the overall cyclostationarity, $X_2$ and $X_5$ assess the $\alpha$-variance, $X_3$ and $X_6$ reflect the $\alpha$-sparsity, $X_7$ describes the power, $X_8$ the $f$-variance of SPD, and $X_9$ the $f$-sparsity of SPD. For better virtual effect, we employ \gls{ccf}, which is a normalized version of \gls{scd}, with
\begin{align}     
	C_{SS}(f,\alpha) &=\frac{D_{SS}(f,\alpha)}{\sqrt{\mathbb{E}\left\{\left\vert S(f+\frac{\alpha}{2})\right\vert^2\right\}\mathbb{E}\left\{\left\vert S(f-\frac{\alpha}{2})\right\vert^2\right\}}}.
\end{align}
The \gls{ccf} of audios with different cyclic patterns and white noise are depicted in Fig.~\ref{fig:ccf_all}. From the \gls{ccf}, we can see that those audio clips are not only the $1^{\mathrm{st}}$ order cyclostationary, but also the $2^{\mathrm{nd}}$ order cyclostationary \cite{bincsc2025}. Additionally, the development of $\Tilde{D}_{SS}(\alpha)$ and $\overline{D}_{SS}(\alpha)$ are depicted in Fig.~\ref{fig:amax_all} and \ref{fig:amean_all}, respectively. The measure $\overline{D}{SS}(\alpha)$ is designed to be sensitive to broadband cyclic patterns, whereas $\tilde{D}{SS}(\alpha)$ is specifically tailored to detect narrowband cyclic components.

 
 \section{Behavioral Measures}\label{sec:survey} 

 We investigate the relationship between extracted statistical features and their actual impact on humans by examining psychological, physical, and relaxing effects. $100$ participants were recruited online for a behavioral survey conducted in March $2024$. To avoid response bias, while the keyword \emph{ASMR} was mentioned during the survey, specific information of \gls{asmr} were omitted.  Specifically, the assessment of psychological effects was conducted by asking participants to rate how strongly they experienced sensations such as brain chills or changes in mental states. Physical effects were evaluated by having participants rate the intensity of sensations like scalp, neck, or back chills. Relaxing effects were evaluated by asking participants to describe whether the clips were relaxing, neutral, or stressful. Both effects were assessed using a $7$-point scale. Although semantic contents were largely removed, which generally created confusion among participants, this does not indicate freedom from cognitive bias. We quantitatively assessed this cognitive bias by asking participants to describe whether the clips seemed natural or artificial to them, also using a $7$-point scale.

 Participants were also required to wear headsets for optimal experience. The survey included seven types of audio clips, each lasting $20$
 seconds: one white noise clip, plus both monophonic and stereophonic clips featuring \gls{fm}, \gls{sr}, and \gls{mir} patterns. Furthermore, we randomize the listening orders of those audio clips.

  \section{Survey results and linear regression with OLS and LMM}\label{sec:results}

  \subsection{Effects of Audio Types}
 
   \begin{figure*} [ht]
    \centering
    \begin{subfigure}[b]{0.325\textwidth}
        \centering
        \includegraphics[width=\textwidth]{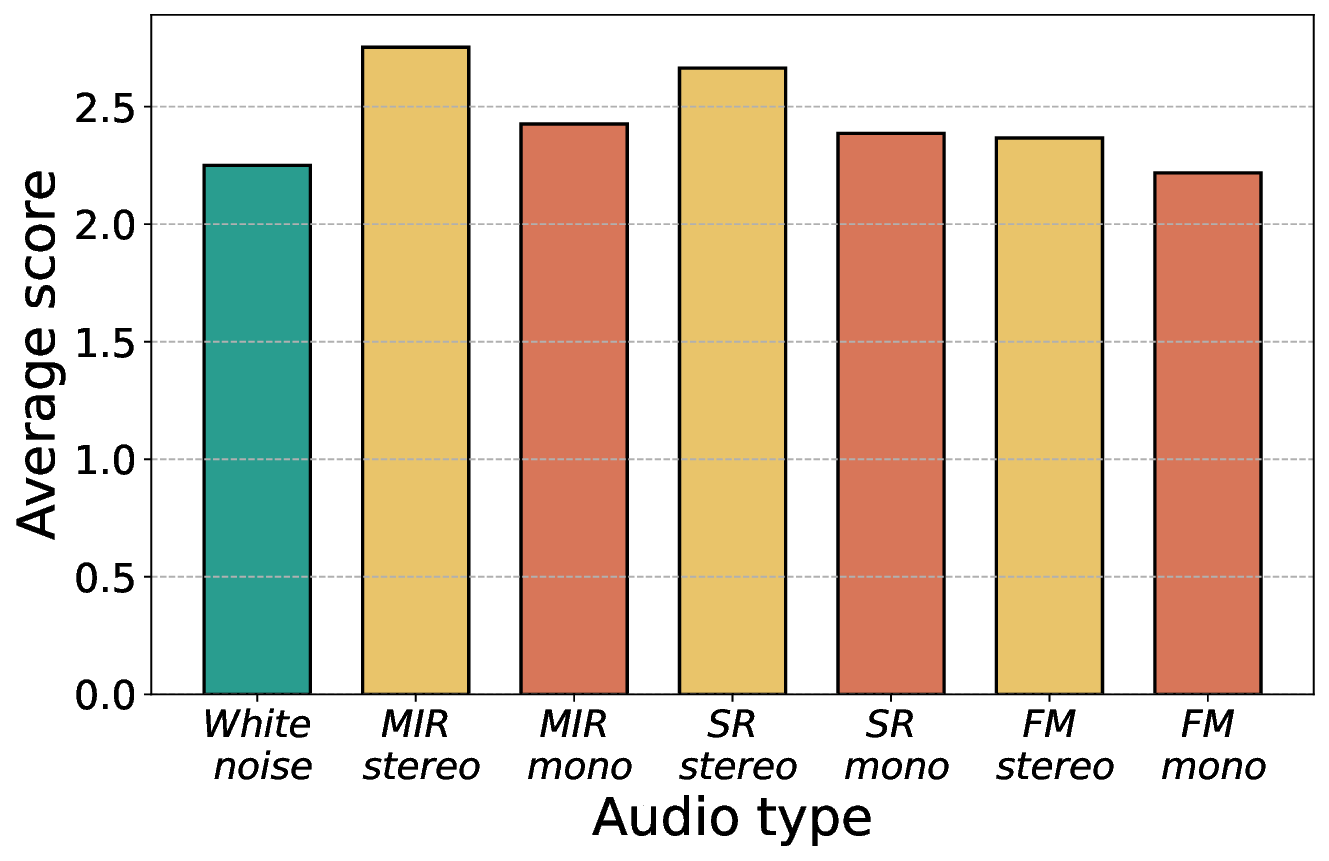}
        \caption{Average physical score per audio type}
        \label{fig:trace_mani_bias}
    \end{subfigure}
    \hfill
    \begin{subfigure}[b]{0.325\textwidth}
        \centering
        \includegraphics[width=\textwidth]{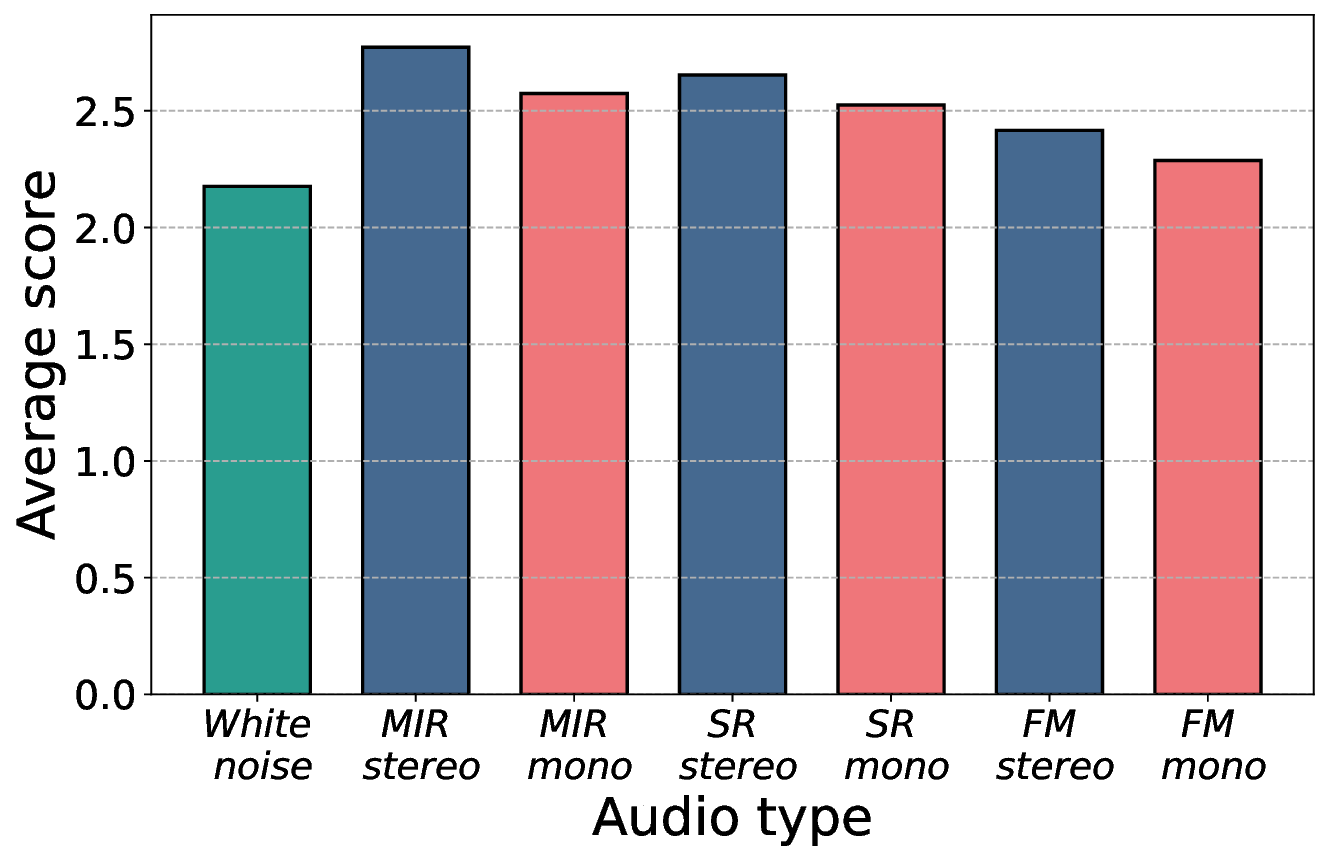}
        \caption{Average psycho. score per audio type}
        \label{fig:trace_mani_ran}
    \end{subfigure}
    \begin{subfigure}[b]{0.325\textwidth}
        \centering
        \includegraphics[width=\textwidth]{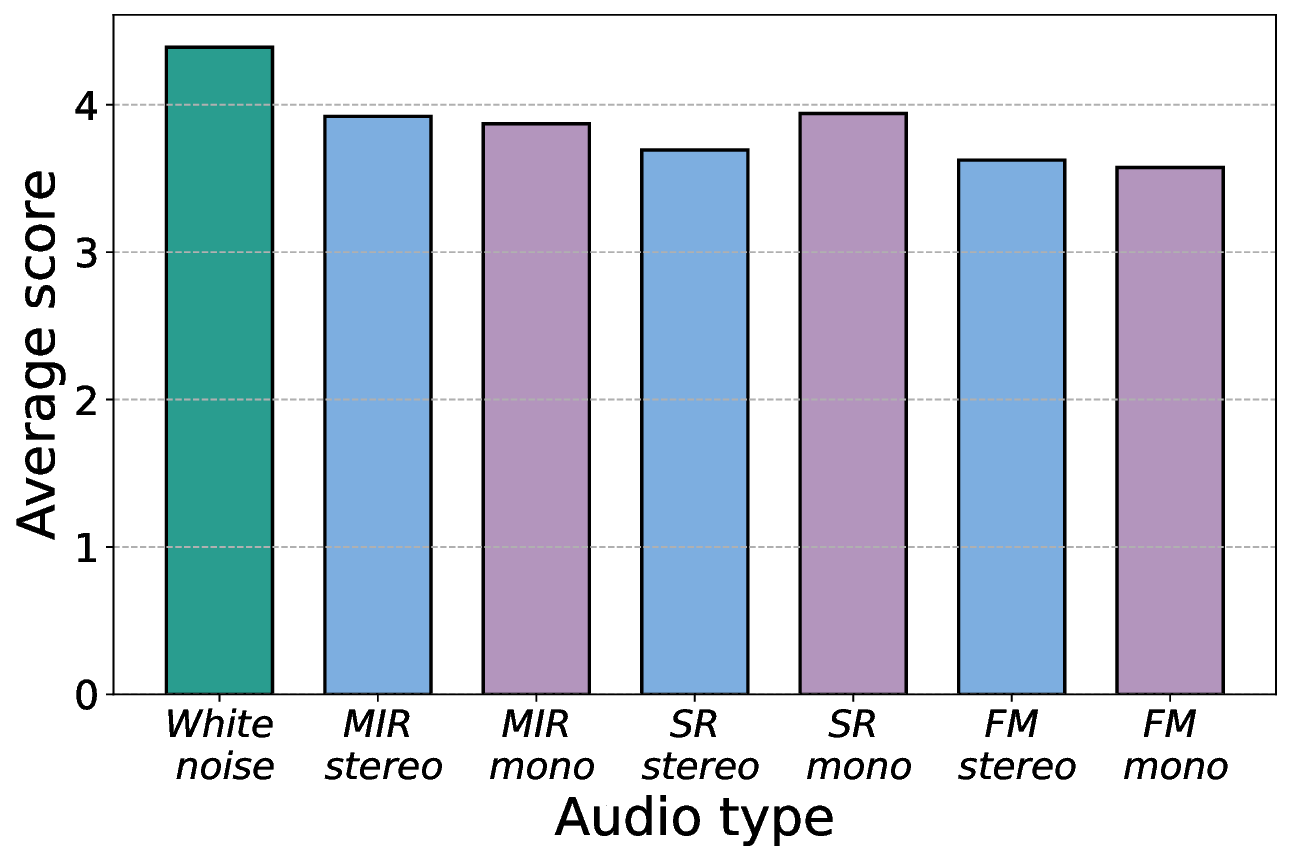}
        \caption{Average relaxing score per audio type}
        \label{fig:trace_bia_coor}
    \end{subfigure}
    \caption{Average effect scores of different audio types}
    \label{fig:average}
   \end{figure*}
   \begin{figure*}[t]
    \centering
    \includegraphics[width=0.85\textwidth]{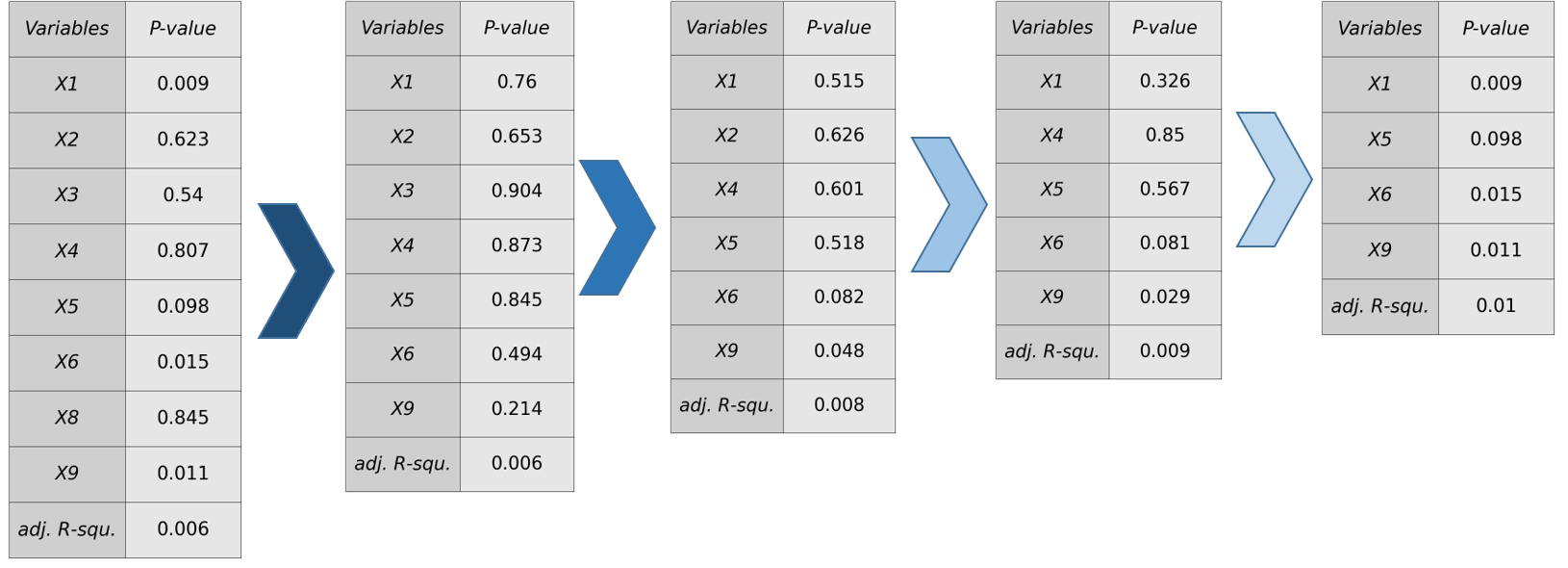} 
    \caption{Model optimization}
    \label{fig:model_opt}
\end{figure*}
  Firstly, we analyze the three aforementioned effects with respect to different audio types. The average scores per audio type are depicted in Fig.~\ref{fig:average}. For both psychological and physical effects, we observe that audio clips with cyclic patterns generally outperform simple white noise, with the score ranking: $\emph{MIR} > \emph{SR} > \emph{FM} > \emph{white noise}$. The results indicate that the more  rigorous the cyclic patterns are, the less likely they are to trigger psychological and physical effects. Additionally, stereophonic effects further enhance such triggers. For relaxing effect, white noise has been observed to be more effective than other audio clips, which subscribes to the \gls{eeg} based analysis in \cite{liasmr2024}. White noise contains minimal informational content, reducing cognitive processing and promoting relaxation. Contrary to our previous findings, stereophonic effects do not contribute to relaxing effects.

  \subsection{Acoustic features and \gls{asmr} response} 
  Secondly, with the statistical features of these \gls{asmr} clips extracted, we can have a more detailed prospective on the impact of audio statistical features on the psychological and physical effects. The regression analysis was conducted using two different models: \gls{ols} and \gls{lmm}. The key difference is that \gls{lmm} considers fixed and random effects on top of the linear relationship, with
  \begin{equation}
  y=\sum\limits_{i=1}^I a_i X_i+\sum\limits_{i=0}^Ib_iz_i+\epsilon,\label{eq:lmm}
  \end{equation}
  where $a_i$ is the coefficient corresponding to the statistical features $X_i$, while $z_i$ represents the random variables (the listening orders) with $b_i$ as their corresponding coefficients. Considering that $X_8$ represents the mean of the spectral, in other words, the volume of the audios, and during the survey we allowed participants to adjust the volume to their most comfortable level, we excluded $X_8$ from our analysis.

  Even though we selected 8-dimensional features, the impact of each feature on \gls{asmr} perception remains unknown. One strategy is to remove the most insignificant variables and redo the regression analysis to determine if the model can be improved, thereby obtaining a more reliable model. As exemplified in Fig.~\ref{fig:model_opt}, our regression results can be improved through variable optimization, with consistent findings across two type of regression. Detailed results of optimized models demonstrated in Tab.~\ref{tab:regression_results}. 
  \subsubsection{ASMR response with consideration of listen orders} 
  In the \gls{lmm}, we considered the listen order as random effects potentially interfering with participants' audio clip processing. For the regression models of psychological and physical effects, the variance in intercepts across listening orders can be neglected, indicating minimal interference with the triggering of effects.
    \begin{table}[h]
    \centering
    \renewcommand{\arraystretch}{1.2}
    \begin{tabular}{lccc}
        \toprule
        \textit{Phy. M. Var.} & \textit{P-value} & \textit{Coef. OLS} & \textit{Coef. LMM} \\
        \midrule
        \textbf{X1} & 0.009  & 2.314   & 2.366  \\
        \textbf{X5} & 0.098  & -138.224 & -139.667 \\
        \textbf{X6} & 0.015  & 25.687  & 26.284 \\
        \textbf{X9} & 0.011  & -8.427  & -8.661 \\
        \toprule
        \textit{Psy. M. Var.} & \textit{P-value} & \textit{Coef. OLS} & \textit{Coef. LMM} \\
        \midrule
        \textbf{X1} & 0.122  & 4.231   & 4.274  \\
        \textbf{X2} & 0.183  & -81.351 & -81.982 \\
        \textbf{X3} & 0.101  & 23.179  & 23.537 \\
        \textbf{X9} & 0.053  & -5.883  & -5.974 \\
        \bottomrule
    \end{tabular}
    \caption{Regression models from physical and psychological effects, respectively.}
    \label{tab:regression_results}
    \end{table}
     \begin{figure}[!htb]
    \centering
    \includegraphics[width=0.345\textwidth]{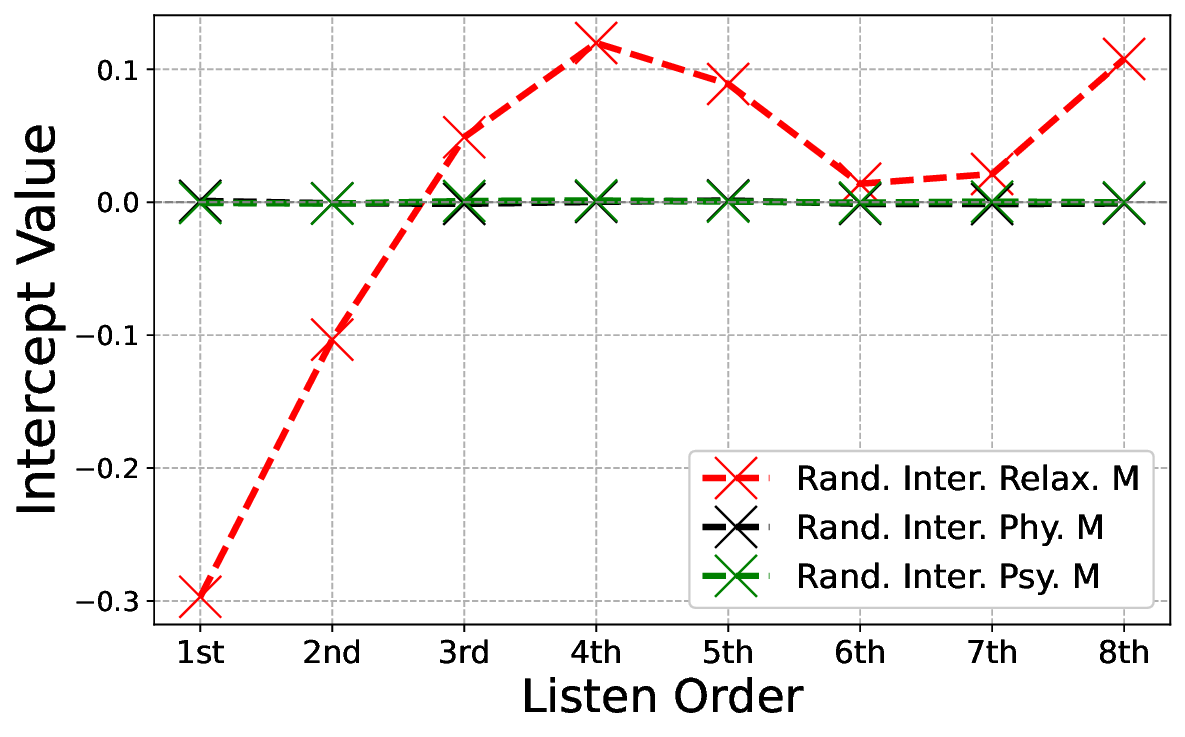} 
    \caption{Intercept of \gls{lmm}}
    \label{fig:intercept}
    \end{figure} 
  
    Conversely, listen order significantly impacts relaxing effects, with the intercept depicted in Fig.~\ref{fig:intercept} revealing a non-monotonically increasing trend. This trend suggests that relaxation perception is a cumulative, time-dependent process where participants' relaxation feeling builds up gradually. On the other hand, the results in Tab.~\ref{tab:regression_results} reveals a positive coefficient for $X_1$ across all regression models, indicating that stronger cyclic patterns intensify observed effects, while the negative coefficient of $X_9$ suggests that smoothly distributed frequency components contribute to triggering these effects. For the regression models of physical effect, the negative intercept for $X_5$ and positive intercept for $X_6$ reveal significant counterbalancing mechanisms. Similarly, for psychological effect, $X_2$ and $X_3$ demonstrate parallel but less significant trends, representing identical descriptors derived from the $\tilde{D}_{SS}(\alpha)$ domain. Overall, these findings suggest that optimal triggering effects require cyclic patterns with a smooth spread but an unequal overall distribution. .
    
    \subsubsection{ASMR response with consideration of cognitive bias}
\begin{figure}[!htb]
    \centering
    \includegraphics[width=0.355\textwidth]{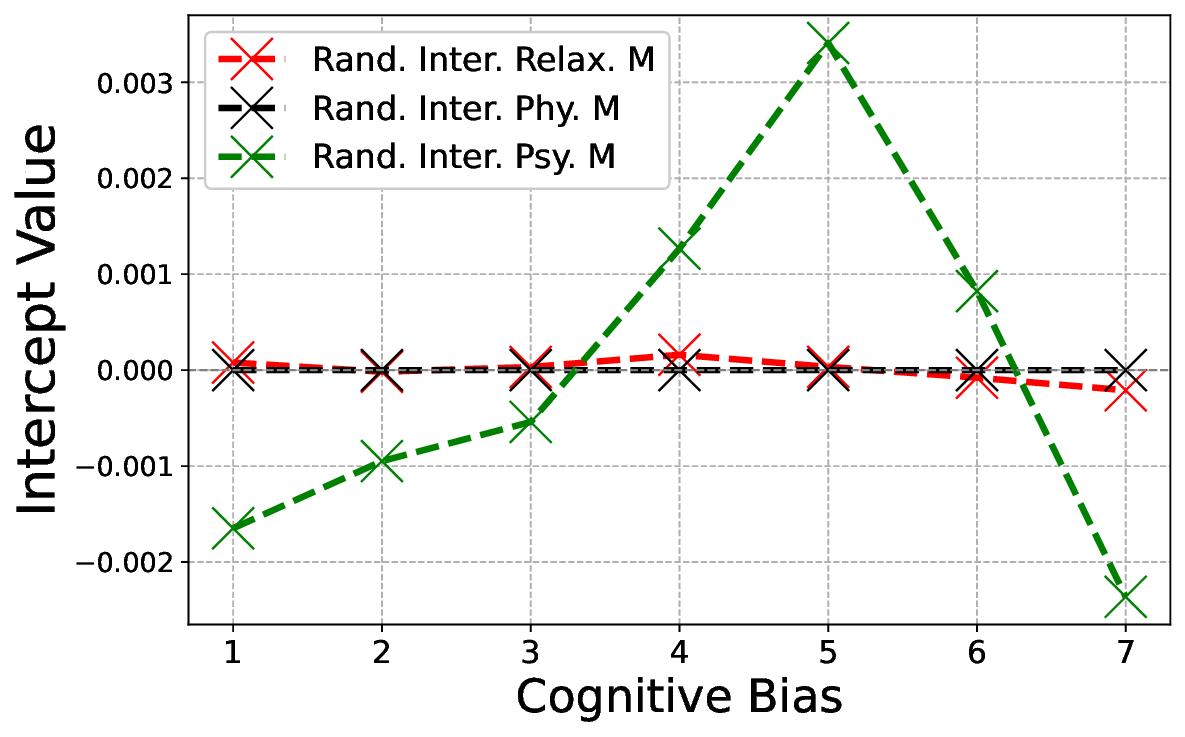} 
    \caption{Intercept of \gls{lmm}}
    \label{fig:intercept2}
    \end{figure}

    In this part, we consider the cognitive bias as random intercepts that potentially impact \gls{asmr} perception. First, we consolidated the average cognitive bias in Tab.~\ref{tab:avgs}, then plotted the intercepts for different models in Fig.~\ref{fig:intercept2} (model coefficients follow identical trends in Tab.~\ref{tab:regression_results}). White noise is regarded as most artificial while \gls{sr} clip is regarded as most natural. The variance of the intercepts follows the order: $\emph{Psy. M} > \emph{Relax. M} > \emph{Phy. M}$, indicating that cognitive bias are most influential for psychological effects in \gls{asmr}, while almost irrelevant for physical effects. Nevertheless, cognitive bias are generally far less impactful compared to listening order, particularly in our setting where semantic content has already been removed. Surprisingly, being natural does not necessarily enhance \gls{asmr} perception; instead, participants with the mindset that the clip is more neutral (cognitive bias = $4$ or $5$) typically experience enhanced psychological and relaxing effects. 
    \begin{table}[h]
    \centering
    \renewcommand{\arraystretch}{1.2}
    \begin{tabular}{lcccc}
         & \textit{White noise} & \textit{MIR} & \textit{SR} & \textit{FM}\\
        \midrule
        \textbf{Sem. Score} & 4.90  & 4.36   & 4.34 & 4.47  \\
        \bottomrule
    \end{tabular}\caption{Average semantic score}\label{tab:avgs}
    \end{table}
    
    \section{Conclusion and Outlook}\label{sec:conclu}
  In this paper, we quantified the relation between \gls{asmr} audio features and human perception across psychological, physical, and relaxing effects. The relaxing effect accumulates progressively and is spatially independent. Cyclic patterns and spatial orientation significantly enhance the psychological and physical effects, both of which are time-invariant. A cognitive bias exists: perceived naturalness modulates the ASMR experience, with neutral ratings correlating with stronger psychological and relaxing effects.

From a signal processing perspective, regression analysis shows that smoothly distributed frequency components, as well as smoothly spread and energy-dense cyclic patterns, yield superior triggering effects. These findings provide quantitative guidelines for engineering \gls{asmr}-inducing acoustic stimuli in human-machine interfaces without relying on semantic content.

\bibliographystyle{IEEEtran}
\bibliography{references}

\end{document}